# Application of Rough Set Theory in Data Mining

Thabet Slimani

*College of Computer Science and Information Technology, Taif University*

**Abstract**

Rough set theory is a new method that deals with vagueness and uncertainty emphasized in decision making. Data mining is a discipline that has an important contribution to data analysis, discovery of new meaningful knowledge, and autonomous decision making. The rough set theory offers a viable approach for decision rule extraction from data.This paper, introduces the fundamental concepts of rough set theory and other aspects of data mining, a discussion of data representation with rough set theory including pairs of attribute-value blocks, information tables reducts, indiscernibility relation and decision tables. Additionally, the rough set approach to lower and upper approximations and certain possible rule sets concepts are introduced. Finally, some description about applications of the data mining system with rough set theory is included.

## I. Introduction

Rough set theory (RST) is a major mathematical method developed by Pawlak in 1982 (Pawlak, 1982). This method has been developed to manage uncertainties from information that presents some inexactitude, incompleteness and noises. When the available information is insufficient to determine the exact value of a given set, lower and upper approximations can be used by rough set for the representation of the concerned set. The approximation synthesis of concepts from the acquired data is the main objective of the rough set analysis. For example, if it is difficult to define a concept in a given knowledge base, rough sets can 'approximate' with respect to that knowledge. In decision making, it has confirmed that rough set methods have a powerful essence in dealing with uncertainties. The RST has been applied in several fields including image processing, data mining, pattern recognition, medical informatics, knowledge discovery and expert systems. In the current literature, several research works have been combined the rough set theory with other artificial intelligence methods such as neural networks, fuzzy logic, additionally to other methods resulting in some good results. The use of rough set theory to solve a specific complex problem has attracted world-wide attention of further research and development, extending the original theory and increasingly widening fields of application. Additionally, rough set as a computationally efficient technique it presents a basic significance to many theoretical developments and practical applications of computing and automation, especially in the areas of machine learning and data mining, decision analysis and intelligent control. Among other computational problems, rough set addresses problems such as data significance evaluation, hidden pattern discovery from data, decision rule generation, data reduction and data-driven inference interpretation (Pawlak, 2004). This paper attempts to offer a concise description of the basic ideas of rough set theory, and its major extensions with data mining applications.

This paper is organized as follows. The next section outlines the theoretical aspects of rough set theory. Section 3 introduces the data representation with RST including the concept of information table, and the concept of decision table. Section 4 describes a range of applications of rough set theories in data mining and as an important topic in automation and computing. The final section concludes the paper.

## II. Rough Set : theoretical aspects

RST can be defined as an extension of the conventional set theory that supports approximations in decision Making (Pawlak, 1982). As described in the introduction, a rough set is the approximation of a vague concept (set) by a pair of fixed concepts that classify the domain of interest into disjoint categories called lower and upper approximations. The description of the domain objects which are known with certainty to belong to the subset of interest is called the lower approximation, whereas the description of the objects which possibly belong to the subset is called the upper approximation.

The fundamental theory of rough sets is described from (Komorowski et al. 1999; Pawlak, 1982). Let be a finite set of objects or universe $\Omega \neq \emptyset$, any subset $A \subseteq \Omega$ of the universe is called a concept in $\Omega$ and each knowledge can be represented by any family of concepts in $\Omega$. The knowledge base over $\Omega$ is referred by the family of classifications over the universe $\Omega$. The consideration of the "universe" to be a finite set reveals The formal foundation of RST. The meaningfulness of updating sets (insert, delete and join operations), for example in database systems is interesting in all database applications (medical database or data warehousing applications). Let be an equivalence relation over $\Omega$ represented by $R \subseteq A \times A$, then the following properties should be taken into account:
1) R is reflexive : aRa,
2) R is symmetric: if aRb then bRa
1) R is transitive (if aRb and bRc then aRc)

U/R is defined as the family of equivalence classes of R and $a_R$ is defined as a category in R that contains an element a included in $\Omega$. Let be a knowledge base KB=($\Omega, R$) if $X \subseteq R$ and $X \neq \emptyset$, then the equivalence relation I(X) is called the indiscernibility relation over X.
The current research in RST explores the complementary mathematical properties with other mathematics disciplines. As an example, the author in (Jarvinen, 2004) has studied the ordered set of rough set theory and showed that the relations are not essentially reflexive, symmetric or transitive. In a similar manner to the definition of $A \subset \Omega$ and $R \in I(KB)$,

1) a=$\underline{R}$A  if and only if $a_R \subseteq A$
2) a=$\overline{R}$A  if and only if $a_R \subseteq A \not\equiv \emptyset$

Where the first inequality is called R-lower approximation of A and the second inequality is called R-upper approximation of A. Additionally, let $P_R(A)=\underline{R}A$ denote the R-positive region of A and $N_R(A)= \Omega - \overline{R}A$ denote the R-negative region of A and $BL_R(A)= \overline{R}A-\underline{R}A$ denote the R-borderline region of A.

The following accuracy measure characterizes the degree of completeness, where $c_R$ depicts the cardinality of the set R as follows:

$$\alpha_R(A) = \frac{c_{\overline{R}}}{c_{\underline{R}}} \quad \text{where } A \neq \emptyset \tag{1}$$

The degree of knowledge completeness is expressed by the above formulated accuracy measure. The previous equation is able to capture the boundary region size of the data sets; but, the structure of the knowledge is not easily captured. However, the fundamental advantage of RST is the ability to hold a category that cannot be piercingly defined given a knowledge base. The Inexactness *and* topological characterization of imprecision can be expressed by the following measures:

1) if $\underline{R}A \neq \emptyset$ and $\overline{R}A \neq \Omega$, then A is roughly R-definable,
2) if $\underline{R}A = \emptyset$ and $\overline{R}A \neq \Omega$, then A is internally R-undefinable,
3) if $\underline{R}A \neq \emptyset$ and $\overline{R}A = \Omega$, then A is externally R-undefinable,
4) if $\underline{R}A = \emptyset$ and $\overline{R}A = \Omega$, then A is totally R-undefinable,

Based on the Eq. (1) and the classifications above rough sets can be characterized by the size of the boundary region and structure.

III. **Data representation with RST**

The paper is based on data-mining-related techniques of the original rough set model. In the following section, some data mining techniques and applications used with RST are reviewed.

RST is based on the assumption that with every object in the universe, we associate some information (data, knowledge). Objects characterized by the same information have a similar view of the available information about them. The similarity relation generated in this way is the mathematical basis of rough set theory.

Elementary set is used to nominate any set of all similar objects, and form atom (basic granule) of knowledge about the universe. The crisp or precise set is used to nominate any union of some elementary sets; otherwise the set is rough (imprecise, vague). The objects included in the available knowledge which cannot be with certainty are classified as members of the set or its complement.

In contrast to precise sets, rough set cannot be characterized in terms of information about their elements. With any rough set approach it is associated a pair of precise sets - called the lower and the upper approximation. All objects which surely belong to the set characterize the lower approximation and all objects which possibly belong to the set characterize the upper approximation.

Data are often presented as a table, where each column is labeled by an *attribute*, each row is labeled by an *object* of interest and each entry of the table contains an *attribute value*. Such tables are composed of *information systems, attribute-value tables* and *information tables*. In this paper we will distinguish data sets in two forms: as information tables and as decision tables. In both cases the columns represent variables and rows represents cases (objects). All variables in information tables are called attributes while in decision tables it is needed to specify one variable called a decision variable, and the remaining variables are attributes.

### A. Information Tables

Information can be represented in a form of a table. Such tables are composed of *information systems, attribute-value tables* and *information tables*. The basic problems that can be tackled by the employment of RST are the following:

- A set of object can be characterized in terms of attribute values.
- It is possible to find total or partial dependencies between objects
- Data reduction
- The more significant attributes can be discovered
- Generation of decision rules

*An example of information table is presented in Table 1*. Three attributes: *Coal*, *Sulfur*, *Phosphorus*. The table contains data concerning six cast iron pipes exposed to high pressure endurance test.

TABLE I
EXAMPLE OF INFORMATION TABLE

| Pipe | Coal | Sulfur | Phosphorus |
|---|---|---|---|
| 1 | High | High | Low |
| 2 | Avg | High | Low |
| 3 | Avg | High | Low |
| 4 | Low | Low | Low |
| 5 | Avg | Low | High |
| 6 | High | Low | High |

Let $\Omega$ represents the set of all cases, the set of all attributes denoted by $A$, and the set of all attribute values denoted by $V$. An information table defines an information function I: $\Omega \times A \rightarrow V$. For example, I(1, Coal) = High.

Let be the attribute-value pair $\tau = (a, v)$ where $a \in A$, $v \in V$. The *block* it is denoted by $[\tau]$, which denotes the set of all cases from $\Omega$ where each attribute $a$ has as value $v$. In the association rule approach of data mining, the support measure of an attribute, compute the existence of an attribute in a specified row, then the support of an attribute-value pair is obtained by the cardinality of $[\tau]$ ($|[\tau]|$). For the information table from Table 1, the block and support are defined as follows:

[(Coal, High)] = {1, 6}, and support([(Coal, High)])=2
[(Coal, Avg)] = {2,3,5}, and support([(Coal, Avg)])=3
[(Coal,Low)] = {4}, and support([(Coal,Low)])=1
[(Sulfur, High)] = {1, 2, 3}, and support ([(Sulfur, High)])=3
[(Sulfur, High)] = {4,5, 6}, and support ([(Sulfur, High)])=3
[(Phosphorus, Low)] = {1, 2,3,4}, and support ([(Phosphorus, Low)])=3
[(Phosphorus, High)] = {5, 6}, and support([(Phosphorus, High)])=2

If we have $x \in \Omega$ and $B \subseteq A$. We denote the elementary set of B containing $x$ by $[x]_B$, represented by the following set: $\cap\{[(a,v)] \mid a \in B, I(x,a) = v\}$

Let be the subset of $\Omega$ containing all cases from $\Omega$ that are indistinguishable from $x$ while using all attributes from B the elementary sets. Elementary sets are called *information granules* in the terminology of *soft computing*. Element sets are blocks of attribute-value pairs represented by that specific attribute, While subset B is limited to a single attribute,. Consequently,

$[1]_{\{Coal\}}=[6]_{\{Coal\}}=[(Coal, High)]=\{1,6\}$
$[2]_{\{Coal\}}=[3]_{\{Coal\}}=[5]_{\{Coal\}} =[(Coal, Avg)]=\{2,3,5\}$
$[4]_{\{Coal\}}=[(Coal, Low)]=\{4\}$

To combine two attribute-values Coal and Sulfur, for example, the elementary set of B=(Coal, sulfur) is defined as follows:

$[1]_B=[(Coal, High)] \cap [(Sulfur, High)]=\{1\}$, and support $([1]_B)=1$
$[2]_B=[(Coal,Avg)] \cap [(Sulfur,High)]=\{2,3\}$, and support $([2]_B)=2$
$[3]_B=[(Coal,Low)] \cap [(Sulfur,Low)]=\{4\}$, and support $([3]_B)=1$
$[4]_B=[(Coal,Avg)] \cap [(Sulfur,Low)]=\{5\}$, and support $([4]_B)=1$
$[5]_B=[(Coal,High)] \cap [(Sulfur,Low)]=\{6\}$, and support $([5]_B)=1$

As another definition, elementary sets may be defined through the notion of an indiscernibility relation. Let B be a nonempty subset of A of all attributes ($B \subseteq A$), the binary relation on $\Omega$ represented by the *indiscernibility relation* IND (B) defined for $x, y \in \Omega$ as follows:

$(x, y) \in$ IND$(B)$ if and only if I$(x, a)$ = I$(y, a)$ for all $a \in B$.

IND($B$) is defined as an equivalence relation. The partitions are a convenient way to present equivalence relations. The $\Omega$ partition is a family of mutually disjoint nonempty subsets of $\Omega$, called *blocks*, such that the union of all blocks is $\Omega$. The partition relative to IND($B$) will be denoted by $P_{\{B\}}$. Each elementary set associated with B represents a block of $P_{\{B\}}$. For example,

$P_{\{Coal\}}=\{\{1, 6\},\{2,3\},\{4\},\{5\}\}$ ➔ support($P_{\{Coal\}}$ )=\{2,2,1,1\}
$P_{\{Coal, Sulfur\}}=\{\{1\},\{2,3\},\{4\},\{5\},\{6\}\}$ ➔ support($P_{\{Coal, Sulfur\}}$)=\{1,2,1,1,1\}

The important subsets of attributes are called reducts. A subset B of the set A is named as a *reduct* if and only if it includes the following properties:

1. $P_{\{B\}} = P_{\{A\}}$ and

2. *B* is reduced with this property, i.e., $P_{(B-\{a\})} \neq P_{\{A\}}$ for all $a \in B$.

As an example, {Coal} is not a reduct since $P_{\{Coal\}} = \{\{1, 3, 4\}, \{2\}, \{5, 6\}\} \neq P_{\{A\}} = \{\{1\}, \{2\}, \{3\}, \{4\}, \{5\}, \{6\}\}$. In a similar manner, {Coal, Sulfur} is not a reduct given that: $P_{\{Coal, Sulfur\}} = \{\{1\},\{2,3\},\{4\},\{5\},\{6\}\} \neq P_{\{A\}} = \{\{1\}, \{2\},\{3\}, \{4\}, \{5\}, \{6\}\}$.

The reducts computation is systematic and based on first checking of all single attributes. As a next step is to check all subsets *B* of *A* with $|B| = n$ {n=2 in our example} such *B* is not a superset of any existing reduct, where $|B|$ is the cardinality of the set *B*.
    The following subsets are reducts:
{Coal, Sulfur} since $P_{\{Coal, Sulfur\}} = P_{\{A\}}$
{Coal, Phosphorus} since $P_{\{Coal, Phosphorus\}} = P_{\{A\}}$
{Sulfur, Coal} since $P_{\{Sulfur, Coal\}} = P_{\{A\}}$

*B. Decision Tables*

A formal definition of a decision table is presented based on the concepts described in (Pawlak, 1982). A decision table is a system S= ( $\Omega$, A, V, f) where:
- $\Omega$ represents the same definition of the universe of the above section
- A denotes the same definition of the set of attributes described in the above section. A is constituted by the union of the conditions attributes set (Cond) and the decision attributes (Dec) (Cond∪Dec)
- V denotes the union of the set of values of an attribute a included in A (domain of a) represented as follows:

$$\bigcup_{a \in A} V_a$$

- f is a total function f:$\Omega$xA→V called the decision function, such that f(x,a) belongs to $V_a$ for every x that belongs to $\Omega$ and every value of an attribute that belongs to A. The decision rule in S is denoted by a function $f_x$:A→V, such that $f_x(a)=f(x,a)$ for every x that belongs to $\Omega$ and every value of an attribute that belongs to A.

An example of decision table is given in Table 2, where attributes are: {Coal, Sulfur, phosphorus} and decision attribute is {Dec}.

TABLE I
EXAMPLE OF DECISION TABLE

| | Attributes Cond | | | Dec |
|---|---|---|---|---|
| *Pipe* | *Coal* | *Sulfur* | *Phosphorus* | *Cracks* |
| 1 | High | High | Low | Yes |
| 2 | Avg | High | Low | No |
| 3 | High | High | Low | Yes |
| 4 | Low | Low | Low | No |
| 5 | Avg | Low | High | Yes |

| 6 | High | Low | High | Yes |

Table 2 encloses data concerning six cast iron pipes exposed to high pressure endurance test. In Table 2, the condition attributes displays the percentage content in the pig-iron of coal, sulfur and phosphorus respectively, whereas the condition attribute *Cracks* revels the result of the test. In the decision table $\Omega=\{1,2,\ldots,6\}$, Cond={Coal, sulfur, Phosphorus}, Dec={Cracks} and domain of all attributes are equal V={High, Avg, Low, Yes, No}.

In RST the approximations determine the dependency (total or partial) between condition and decision attributes. Given that the quality of pipes cannot be determined exactly by the content of coal, sulfur and phosphorus in the pig-iron, it is possible to use approximations to identify the quality of the pipes (identifying the functional relationship between values of condition and decision attributes.

The consistency factor (conflicting decision rule number of all decision rules in the table) of the decision table may define the degree of dependency between condition and decision attributes. The use of decision rules conflicting means the use of rules having the same conditions but different decisions. As an example, the Table 2 consistency factor is 4/6. Consequently, this factor means that four out of six (ca. 60%) pipes can be appropriately classified as good or not good on the basis of their composition.

Let B a subset of A. It is possible to assign to every subset X of the universe $\Omega$ two sets $\underline{B(X)}$ and $\overline{B(X)}$ called, respectively, the *B-lower* and the *B-upper approximation* of X specified as follows:

- $\underline{B(X)} = \{x \in \Omega: B(x) \subseteq X\}$
- $\overline{B(X)} = \{x \in \Omega: B(x) \sqcap X \neq \emptyset\}$

The following coefficient of rough set can characterize the accuracy of the approximation:

$$a_B(X) = \frac{|\underline{B(X)}|}{|\overline{B(X)}|}$$

Where $|\underline{B(X)}|$ denotes the cardinality of $\underline{B(X)}$. It is clearly that $0 \leq a_B(X) \leq 1$. If $a_B(X)=1$, X is crisp with respect to B (X is precise with respect to B), and otherwise, if $a_B(X) < 1$, X is rough with respect to B. It is possible to use approximation to define total or partial dependencies between attributes, decision rule generation, reduction of attributes and others, but will not discuss these issues here.

A decision rule $f_x$ is included in $\Omega$ is consistent or deterministic if for every y included in $\Omega$, y≠x ($f_x$/Cond= $f_y$ /Cond)➔ ($f_x$/Dec= $f_y$ /Dec); otherwise the decision rule $f_x$ is nondeterministic or inconsistent. In a similar manner, a decision table S is deterministic if all of its decision rules are deterministic; otherwise the decision table S is nondeterministic.

Let be R1 and R2 an example of decision rules as follows:
**R1**:IF (Coal=High AND Sulfur=High AND Phosphorus=Low) Then (Crack=Yes)
**R2**: If (Coal=Avg) Then (Crack=Yes) OR (Crack=No)
A decision rule may be characterized by the most specific definitions as follows:

- *Rule length* is the elementary condition element number of the rule. As an example, the length of R1=3.
- *Rule strength* is count of objects in the data set having the property described by the rule conditions and decisions. As an example, the rule strength of R1=4.
- *Exact rule* : the outcome of an exact rule corresponds to one or more different conditions. Exact rules are generated from the set of objects in the lower approximation. As an example, R1 is an exact Rule.
- *Approximate rule:* The same condition of an approximate rule corresponds to more than one outcome. Approximate rules are generated for the boundary. R2 is an example of approximate rule.
- *Rule support:* is the count of all objects in the data set having the property described by the conditions of the rule. As an example, the rule support of R1=3.
- *Rule coverage*: is the proportion of objects contained in the training set, identified by this rule. As an example, the rule coverage of R1=3/4=0.75.
- *Rule acceptance*: the rule acceptance measure may be expressed as the count of condition terms of a rule. It is a subjective measure that reflects the confidence of the user in the extracted rules. It is a generalization of the rule support and rule coverage.
- *Discrimination level (DL):* DL measures the level of precision of a rule that represents the corresponding objects.
- *Decision support measure* (DSM): is the total number of rules that support a decision. A DSM may be expressed by the number of objects from the training set supporting the decision.
- *Decision redundancy factor* (DRF): is the count of mutually exclusive feature sets related to the same decision.

## IV. Application of RST in Data Mining

RST has found a lot of interesting applications. Particularly, it appears of particular importance to decision support systems and data mining. The propriety of the absence of any preliminary or additional information about data is the main advantage of RST. The application of rough set theory is successful in many real-life problems like engineering, banking, medicine, pharmacology, financial and market analysis and others. A number of perfect applications are listed in the following section:

- **Rough set approach to materials science**: this approach provides a new algorithmic method for predicting and understanding material properties and behavior, which can be very useful in creating new materials (Jackson et al., 1996). Rough sets to material sciences are firstly applied in (Jackson et al., 1994; Jackson et al., 1996), which presents a great interest to this community.
- **Rough set Applications requiring suitable software**: A lot of software systems for computers based on RST have been developed. The most known include Rough DAS and Rough Class (S³owiñski, 1992), LERS (Grzymala-Busse, 1992), and DATALOGIC (Tsumoto, 1996). Some of that software is commercial.
- **LERS Software:** The first version of LERS was developed in 1988 at the University of Kansas. Currently, LERS version is essentially a family of data mining systems. The LERS main objective is to compute decision rules from data. The classification of new cases or the interpretation of knowledge is based on the computed rule sets. The rule computation of LERS system starts from imperfect data (Grzymala-Busse, 1992) (e.g., data characterized by missing attribute values or inconsistent cases). To deal with numerical attribute, LERS also uses a set of discretization schemas. In addition, LERS includes a variety of methods helping to handle missing attribute values. LERS computes lower and upper approximations of all set involved, for inconsistent data (that belongs to two different target sets and characterized by the same values of all attributes). LERS system was used in other areas (e.g., in the medical field by the comparison of the effects of warming devices for postoperative patients, assessing preterm birth) (Woolery &Grzymala-Busse, 1994) and used to diagnoses the melanoma (Grzymala-Busse et al., 2001).

- ☒ **Other applications**: Some different applications of rough set theory can be found in (Lin &Wildberger, 1995; Lin & Cercone, 1997; Sowiñski, 1992; Tsumoto et al., 1996; Wang, 1995; Ziarko, 1993; Wang, 1997). Particularly, some sources realized experiments based on RST for pattern recognition, including speech recognition, music fragment classification, and handwriting recognition, medical diagnosis and control (Brindle & Ziarko, 1999; Kostek, 1998; Plonka & Mrozek, 1995; Shang & Ziarko, 2003; Peters et al., 1999; Mrozek, 1986). These technologies indicate that the trend to develop applications based on extensions of RST will continue.

V. **Conclusions**

The applications of data mining based on the original approach of rough set theory, have been attempted valuable methods to generate decision rules in recent years (about 20 years now). The obtained results need more research, particularly, when quantitative attributes are involved. Due to space limits, this article states only data representation with rough set theory (information and decision table that deal with consistent data), in addition to some limited applications of data mining. As an example of successful rough set theory application of data mining, this paper states the LERS data mining system. As further work, we plan to extend the data representation to be applied to inconsistent data.